# Universal Scaling in the Branching of the Tree of Life

E. Alejandro Herrada[1], Claudio J. Tessone[1,2], Konstantin Klemm[3], Víctor M. Eguíluz[1]*, Emilio Hernández-García[1], Carlos M. Duarte[4]

1 IFISC, Instituto de Física Interdisciplinar y Sistemas Complejos (CSIC-UIB), Palma de Mallorca, Spain, 2 ETH Zürich, Zürich, Switzerland, 3 Bioinformatics Group, Department of Computer Science, University of Leipzig, Leipzig, Germany, 4 IMEDEA, Instituto Mediterráneo de Estudios Avanzados (CSIC-UIB), Esporles, Spain

**Abstract**

Understanding the patterns and processes of diversification of life in the planet is a key challenge of science. The Tree of Life represents such diversification processes through the evolutionary relationships among the different taxa, and can be extended down to intra-specific relationships. Here we examine the topological properties of a large set of interspecific and intraspecific phylogenies and show that the branching patterns follow allometric rules conserved across the different levels in the Tree of Life, all significantly departing from those expected from the standard null models. The finding of non-random universal patterns of phylogenetic differentiation suggests that similar evolutionary forces drive diversification across the broad range of scales, from macro-evolutionary to micro-evolutionary processes, shaping the diversity of life on the planet.

**Funding:** We acknowledge financial support from MEC (Spain) and FEDER, project FISICOS, from CSIC (Spain) project PIE 200750I016, from SBF (Switzerland) through project C05.0148 (Physics of Risk), and from the European Commission through the NEST-Complexity project EDEN. The funders had no role in study design, data collection and analysis, decision to publish, or preparation of the manuscript.

**Competing Interests:** The authors have declared that no competing interests exist.

* E-mail: victor@ifisc.uib-csic.es

## Introduction

The Tree of Life is a synoptic depiction of the pathways of evolutionary differentiation between Earth life forms [1], and contains valuable clues on the key issue of understanding the diversification of life in the planet [2]. The branching pattern of the Tree of Life, which is being captured at increasing resolution by the advent of molecular tools [3], can be examined to investigate fundamental questions, such as whether it follows universal rules, and at what extent random differentiation mechanisms explain the shape of phylogenetic trees. The examination of the structure of the Tree of Life can also help to infer whether evolution acts at intraspecific scales in a way different from the action of evolution at the interspecific scale. Here we address these fundamental questions on the basis of a comprehensive comparative analysis of phylogenetic trees representing different fractions and domains of the Tree of Life, from interspecific to intraspecific scales. We draw from previous analyses of the geometry of the Tree of Life [4], the characterization of other branching systems [5,6], and using tools derived from modern network theory [7–10] to examine the scaling of the branching in the Tree of Life [11,12]. Our analysis is based on a thorough data set of more than 5000 interspecific phylogenies and a sample of 67 intraspecific phylogenies (see Text S1), thereby testing the universality of the results derived across scales.

A phylogenetic tree is a set of nodes, each node representing a diversification event, connected by branches (links). For each node $i$, a subtree $S_i$ is made up of a root at node $i$ and all the descendant nodes stemming from this root. The subtree size $A_i$ gives the number of subtaxa that diversify from node $i$ (including itself). Beyond this measure of the diversity degree, the characterization of how the diversity is arranged through the phylogenies can be achieved through the cumulative branch size, $C_i$, a measure of the subtree shape. It is defined [13] as the sum of the branch sizes associated to all the nodes in the subtree $S_i$, $C_i = \Sigma A_j$. For the same tree size, and restricting to binary branching events, the smallest value of the cumulative branch size is obtained for a completely symmetric, balanced tree, whereas the most asymmetric, the pectinate or comb-like tree in which all branches split successively from a single one, yields the largest $C_i$ value [13]. To be clearer, we show in Figure 1 the analysis of $A_i$ and $C_i$ for a completely balanced tree (Figure 1A) and for a completely imbalanced tree (Figure 1B). A portion of a real phylogenetic tree is also shown (Figure 1C). How the shape of the tree (i.e., the distribution of the biological diversification) does change with tree size (i.e., with the number of taxa it contains) is given by the scaling of the subtree shape $C$ vs. the subtree size $A$, as described by the allometric scaling relation $C \sim A^\eta$. We quantitatively characterize the shape of each tree in our data set by calculating the functions $F(A)$ and $F(C)$, which are the complementary cumulative distribution functions (CCDF) of $A_i$ and $C_i$ values in the tree, respectively, and the value of the allometric scaling exponent, $\eta$. We compare the results derived from the analyses of inter- and intra-specific phylogenetic trees among them, to test for the preservation of branching patterns across evolutionary scales, and against those derived from the analyses of randomly-generated trees to test whether the allometric scaling derived can be modeled using simple, random branching rules.

## Results

The branch-size CCDF displays power-law tails of the form $F(A) \sim A^{1-\tau_A}$ for large branch size $A$ (Figure 2A). The power-law exponents $\tau_A$ are remarkably similar for the data sets analyzed: $\tau_A = 1.76 \pm 0.03$, and $1.74 \pm 0.02$ for intra- and interspecific phylog-enies, respectively. Similarly, the cumulative-branch-size CCDF also displays a power-law tail of the form $F(C) \sim C^{1-\tau_C}$ at large $C$, with a similar agreement between the exponents of the intra- and interspecific data sets: $\tau_C = 1.53 \pm 0.02$ and $1.53 \pm 0.02$, respectively (Figure 2B). The discrepancy observed between the two data sets at the tail of the distributions can be explained by the different sizes of the typical trees on them: each tree contributes a natural cutoff to the overall distribution, and since the intraspecific trees are smaller in average, their cutoff appears at smaller tree sizes.

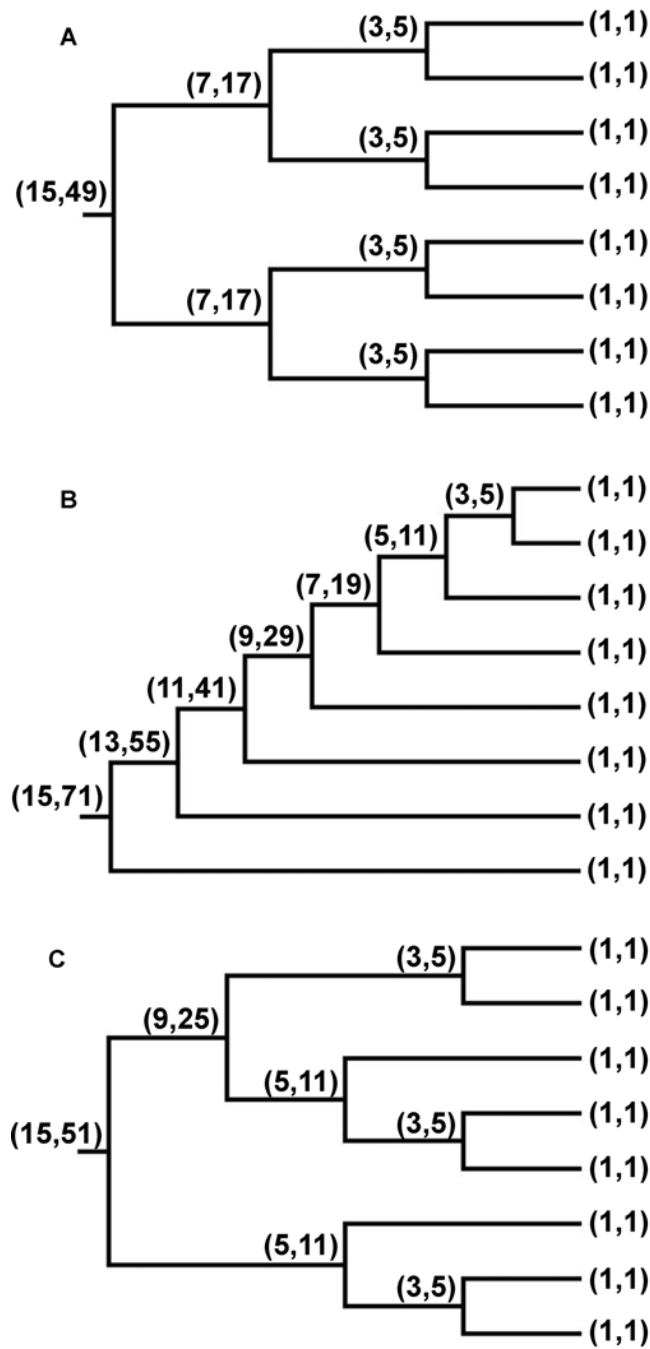

**Figure 1. Branch size and cumulative branch size examples.** The values of the branch size (A) and of the cumulative branch size (C) are shown (in brackets, as (A,C)) at each node of three small example trees. A: a completely balanced tree of 15 nodes; B: a completely imbalanced tree of 15 nodes; C: a subtree of 15 nodes of a real phylogenetic tree, the intraspecific *Vibrio vulnificus* phylogeny presented in full in Fig. S2A. Note that the value of C at the root is maximum for the fully imbalanced tree, and minimum for the balanced one.
doi:10.1371/journal.pone.0002757.g001

The allometric exponent, $\eta$, that characterizes the scaling of tree shape with tree size (Figure 3A), is also remarkably similar for the intraspecific ($\eta = 1.43 \pm 0.01$) and the interspecific ($\eta = 1.44 \pm 0.01$) phylogenies. This constancy of the exponents is still more remarkable when realizing (inset of Figure 3A) that it does not only apply to average properties of sets of intraspecific and interspecific trees, but also to individual phylogenies of groups of organisms pertaining to different kingdoms and living across widely contrasting environments, as it is reflected by the very narrow range of $\eta$ obtained from different phylogenies ($\langle\eta\rangle = 1.47$, $\sigma = 0.03$, Figure 3A). The scaling exponents for our large interspecific data set are also matched almost perfectly (Figure S1) by those derived from a set of 67 interspecific phylogenies randomly drawn from the published literature thereby validating the uniformity of the scaling rules of the broad interspecific phylogenies and the smaller set of intraspecific ones used here. The later was also derived from a similar random sample taken from the published literature (see Text S1).

The allometric scaling of $C \sim A^{1.44}$ derived from our analysis falls somehow in between those obtained by simulated phylogenies derived from two extreme topologies: The symmetric tree gives $C \sim A \ln A$, which corresponds to $\eta = 1$ with a logarithmic correction, while the pectinate tree has $\eta = 2$. The natural null model for tree construction, the Equal-Rates Markov (ERM) model [14,15], yields a scaling $C \sim A \ln A$ similar to the symmetric tree with $\eta = 1$ but different from the scaling displayed by empirical inter- and intraspecific phylogenies, particularly for large ones (Figure 3B). Therefore some topological aspects of phylogenetic trees are not adequately reproduced by the ERM model. Our results imply that successful lineages diversify more profusely than expected under random branching, generating the large imbalances that characterize emerging depictions of the Tree of Life [4]. Alternative models introducing correlations, such as the proportional-to-distinguishable-arrangements (PDA) model [4,16] or the beta splitting model [17], could generate more realistic phylogenies. Guided by previous biological allometric scaling analysis, we have assumed a power-law scaling of the form $C \sim A^\eta$. However, other ansatz could also fit the data. The important point, however, is that these modeling approaches should give similar scaling properties for intra- as for interspecific branching.

## Discussion

Traditionally, microevolutionary and macroevolutionary processes have been studied independently by population geneticists and evolutionary biologists, respectively [18]. The divide between these two levels of generation of biological diversity is an old one, rooted in the controversy between Darwinian gradualism and the saltationism proposed by others, prominently paleontologists, to explain macroevolutionary processes [19]. The debate as to whether macroevolution is more than the accumulation of microevolutionary events remains active [18,20,21], although refined paleontological evidence supports the continuum between micro- and macroevolution for some lineages [22]. The results presented here show that the branching and scaling patterns in intraspecific and interspecific phylogenies do not differ significantly for the topological properties we have calculated. Thus, shall saltation processes be a factor at the macroevolutive level, this is not reflected in the topology of phylogenetic branching as examined here. Evidence for possible differences in phylogenetic topologies between the inter- and intraspecific levels may require a detailed analysis of branching times, which we have not attempted.

PLoS ONE | www.plosone.org    2    July 2008 | Volume 3 | Issue 7 | e2757

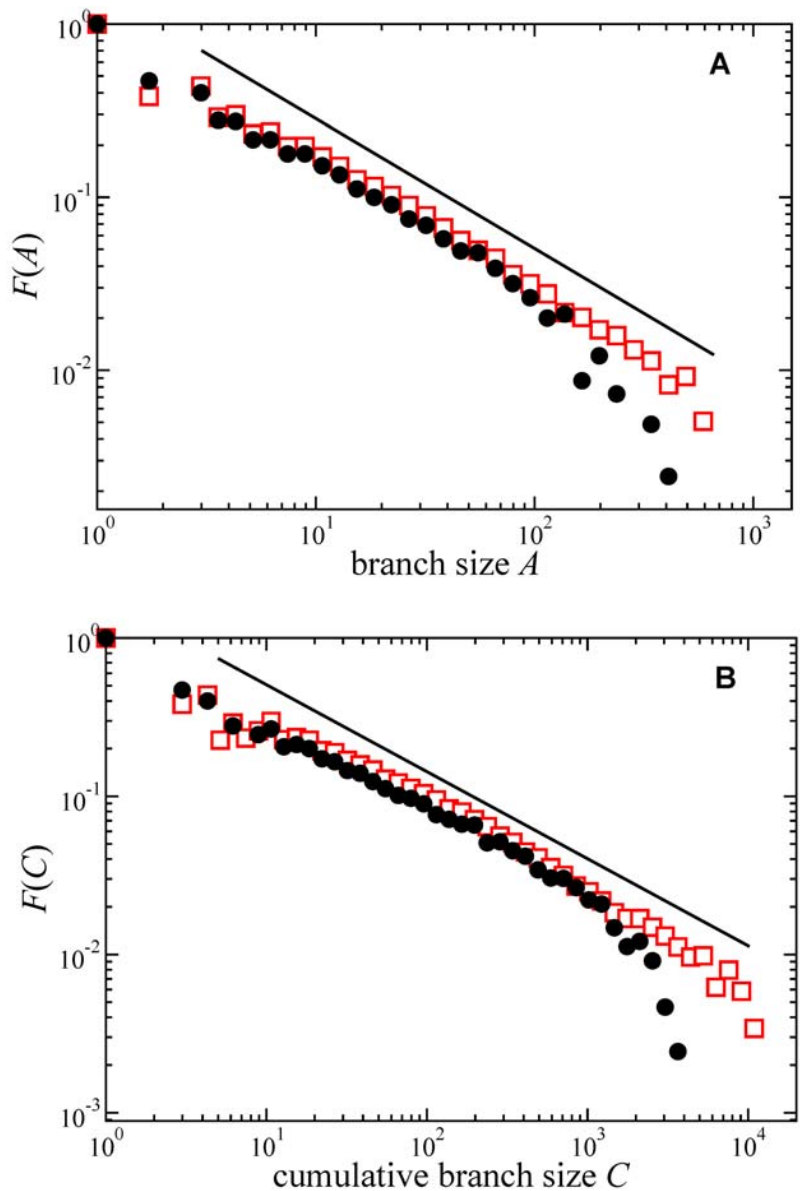

**Figure 2. Average distributions.** Cumulative complementary distribution functions (CCDFs) averaged and logarithmically binned over all phylogenetic trees in the interspecific (empty squares) and intraspecific (solid circles) data sets. A: CCDF of branch size, $F(A)$. Solid line corresponds to a power law $F(A) \sim A^{1-\tau_A}$ with the exponent given by the best fit to the interspecific data set $\tau_A = 1.74$. B: CCDF of the cumulative branch size, $F(C)$. The line corresponds to a power law with the exponent given by the best fit to the interspecific data set $\tau_C = 1.53$.
doi:10.1371/journal.pone.0002757.g002

Processes leading to scaling laws in size distributions in natural systems have been formulated as growth models [23,24]. Many of the findings carry over to scaling properties found in networks [25] and their description in terms of branching processes [26]. But most of these models predict branching topologies similar to the ERM model. An alternative approach to understand the observed exponent would be to trace analogies with scaling laws in different branching systems [5,6,27] which have been explained by invoking a natural optimization criterion based in the fact that the observed trees contain the largest possible number of apices within the smallest number of branching levels. For binary trees of size $A$, where nodes are restricted to occupy uniformly a $D$ dimensional Euclidean space, the minimum value of $C$ scales as $A^\eta$, with $\eta = (D+1)/D$. This scaling also describes the $D$-dimensional tree with the maximum size for a given depth (the average distance between root and leaves). The value of $\eta$ obtained in our phylogeny analysis, $\eta \cong 1.44$, is achieved only for optimal trees restricted to spaces of $D \cong 2.27$ dimensions. Given the apparently unlimited number of variables that may yield differences among taxa, restricting their representation to a space with such a small number of dimensions seems unreasonable. This interpretation suggests that the evolutionary process yielding the observed phylogenies is not the most parsimonious one, which could potentially yield a similar biodiversity with fewer branching levels. In fact, the natural choice $D = \infty$ gives an optimal exponent $\eta = 1$, which correspond to the ERM value and departs from observed scaling. Optimal traffic networks [28] also led to the exponent $\tau_A = 2$ which departs from the empirical scaling exponent reported here for phylogenetic trees.

In summary, the remarkably similar allometric exponents reported here to characterize universally the scaling properties of

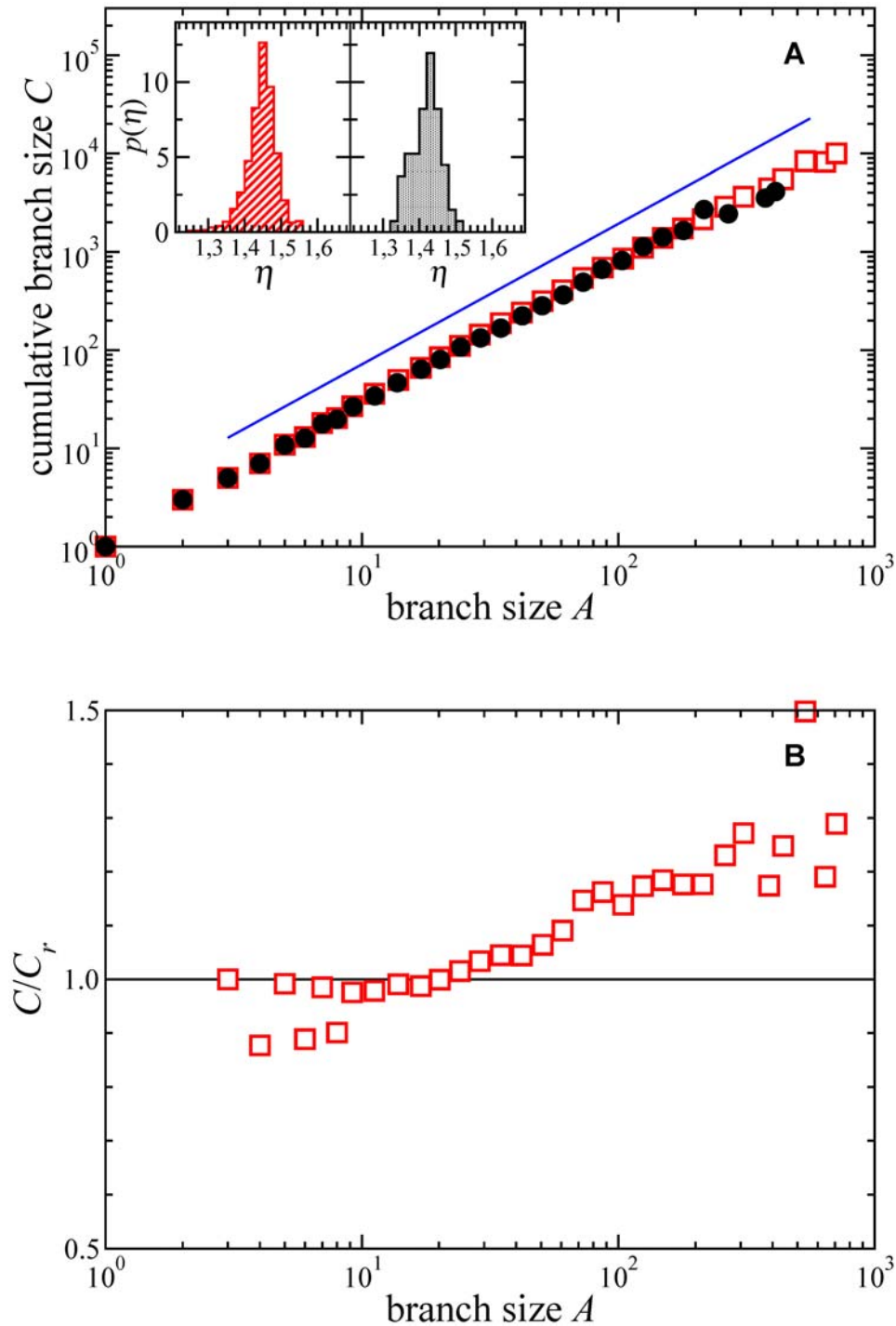

**Figure 3. Allometric scaling.** A: Plot of the logarithmically binned set of values of branch size, $A$, and cumulative branch size, $C$, for the interspecific (empty squares) and intraspecific (solid circles) data sets considered. The line corresponds to a power law $C \sim A^\eta$, with the exponent given by the best fit through all data, $\eta = 1.44$. The inset shows probability distributions of the values of $\eta$ fitted to each individual tree (left: interspecific, right: intraspecific data sets) illustrating the small dispersion in the values. B: Plot of the logarithmically binned set of values of $C$ as a function of $A$ for the interspecific data, normalized by the prediction from the ERM model (the horizontal line). Data systematically deviate from ERM, especially for large size $A$.
doi:10.1371/journal.pone.0002757.g003

intra- and inter-specific phylogenies across kingdoms, reproductive systems and environments, strongly suggests the conservation of branching rules, and hence of the evolutionary processes that drive biological diversification, across the entire history of life. Although at short branch sizes the topology of observed phylogenies cannot differ much from that expected under random and symmetric trees, due to the restriction of binary bifurcations in phylogenetic tree reconstruction, significant departures become universally evident as trees become larger, where the null ERM model and real phylogenies differ (Figure 2B). These deviations suggest (a) that the evolution of life leads to less biodiversity than an optimal tree can possibly generate; and (b) the operation of a mechanism generating a correlated branching, where some memory of past evolutionary events is maintained along each branch. This correlated branching pattern implies that entities that diversify faster than average lead to new biological forms that diversify more than average themselves. Invariance across the broad scales considered here indicates that relatively simple rules govern the phylogenetic branching and the unfolding of biodiversity. Their deviation from random models indicates that evolutionary success is a correlated trait within lineages, yielding present asymmetries in the structure of the Tree of Life.

## Materials and Methods

### Phylogenies databases

On June 30th 2007 we downloaded the 5,212 phylogenetic trees available at that time in the database TreeBASE (http://www.treebase.org). TreeBASE constitutes a large database of interspecific phylogenies, which were collected from previously published research papers. The size of trees oscillates from 10 to 600 tips. Most of the bifurcations in these trees are binary, as confirmed by the fact that the ratio between the number of tips and the total number of nodes gives 0.52 when averaged over all the trees (for perfect binary trees, the ratio is 0.50).

As a comprehensive database comparable to TreeBASE does not exist for intraspecific phylogenies, we constructed an intraspecific data set by manually compiling 67 intraspecific phylogenies from several published phylogenetic analysis [S1–S45]. We compiled this data set in such a way that it contains: 1) Organisms from the main different environments (terrestrial, marine and fresh water), climatic regions (from polar to desert), and branches of life (Table S1). 2) Phylogenetic trees reconstructed with the main phylogenetic tree estimation methods, i.e., neighbor-joining, maximum parsimony and maximum likelihood methods.

In order to test whether the results derived from the examination of the relatively small (67 phylogenies) intraspecific data base can be compared with the much larger (5212) set of interspecific phylogenies extracted from TreeBASE, we sampled the literature to construct a dataset of 67 interspecific phylogenies drawn from the literature [S46–S85] using the same criteria as those to derived the intraspecific phylogeny data base (Table S1), obtaining full agreement (Figure S1). The intra- and interspecific phylogenies derived from the literature ranged between 30 and 170 tips, and they contained mainly binary branching events. An example for each kind of phylogenies is shown in Figures S2A and S3A.

### Branch size and cumulative branch size distributions

We associate to each node $i$ of a phylogenetic tree two quantities, the size $A_i$ (number of nodes) of the subtree $S_i$ made up of node $i$ and all the descendant nodes below it, that is, the subtree which does not contain the global root of the original tree, and the cumulative branch size, $C_i$, defined as the sum of the branch sizes associated to all the nodes in the subtree $S_i$, $C_i = \Sigma A_j$. To characterize the probability distributions of the $A_i$ and $C_i$ values on a particular phylogenetic tree we compute the respective complementary cumulative distribution functions (CCDF): $F(A) = \text{probability}(A_i > A)$, and $F(C) = \text{probability}(C_i > C)$. We observe that these quantities scale, for large values of $A$ and $C$, as power laws: $F(A) \sim A^{1-\tau_A}$ and $F(C) \sim C^{1-\tau_C}$. The exponents $\tau_A$ and $\tau_C$, thus, characterize the probabilities of $\{A_i\}$ and $\{C_i\}$: $P(A) \sim A^{-\tau_A}$ and $P(C) \sim C^{-\tau_C}$, respectively.

### Allometric scaling relationship

We observe that a functional relationship among the values of $C$ and $A$, i.e. among shape and size, exists and also follows a power law, $C \sim A^\eta$, characterized by an exponent $\eta$. Since this relationship encodes the variation of a system property as size is varied, we can call this an *allometric scaling relationship*, to stress its connections with other functional relationships relating function and size [11,13,27]. We note that introduction of the change of variables $C \sim A^\eta$ into $F(C) \sim C^{1-\tau_C}$ leads to $F(C) \sim A^{\eta(1-\tau_C)}$, from which $\eta = (1-\tau_A)/(1-\tau_C)$. Thus, only two out of the three exponents are independent. As simple examples for which the above exponents can be computed by direct counting, we mention the pectinate or fully unbalanced tree, i.e. a tree in which all branching occurs successively along a single branch, characterized by the exponents $\tau_A = 0$, $\tau_C = 1/2$, $\eta = 2$, or the fully symmetric or Cayley tree, characterized by $\tau_A = 2$, and $C \sim A \ln A$, which except for the weak logarithmic correction corresponds to $\eta = 1$ and $\tau_C = 2$. Figures S2B and S3B show, in contrast, the allometric scaling relationship for the particular examples of intra- and inter-specific phylogenies displayed in Figures S2A and S3A.

In order to investigate whether observations differ from random expectations, we have compared the allometric scaling found here with the prediction of a null model [29], the Equal-rates Markov (ERM) model. The ERM model was attributed to Harding [30], and to Cavalli-Sforza and Edwards [31], although it is based on models of the diversification process that date back at least to Yule [23]. The main assumption of the ERM model is that the phylogeny is the product of random branching. This is the result when the "effective speciation rate" (the difference between extinction and speciation rate) is equal for all species. The effective speciation rate may change chronologically, provided that it is the same for all lineages at a given time [23]. For this model we obtain $C \sim A \ln A$, or $\eta = 1$, and also $\tau_A = \tau_C = 2$. The random asymmetries introduced by the ERM are not strong enough to change the scaling behavior from the symmetric tree result.

The quantity $C_i/A_i$ can be thought as a measure of the average *depth* or *distance* of the phylogenetic tree leaves to the node $i$. This can be seen taking into account that $C_i = \Sigma(d_{ij}+1)$, where $d_{ij}$ corresponds to the distance of each of the nodes $j$ of the subtree $S_i$ to the root $i$. Thus, the relationship between $C$ and $A$ can be written as $C_i = A_i + \langle d \rangle_i A_i$, where $\langle d \rangle_i$ is the average depth of the nodes in the subtree $S_i$. The relationship between $C_i/A_i$ and the depth is obtained: $C_i/A_i = \langle d \rangle_i + 1$. This quantity is closely related to the Sackin's index defined as the distance of the leaves to the root: $S = \Sigma_{l \in leaves} d_{l,root}$ [32,33]. It can be shown that for binary trees $C = 2S+1$, where $C = \Sigma_{\forall i} d_{i,root}$. Since the scaling law relating the increase of the depth or Sackin's index with three size is known to be the same as the scaling of the Colless' index, measuring the symmetry or balance of a phylogenetic tree [34], our results for $\eta$ can be put in the context of the numerous studies available on the unbalance of phylogenetic trees [4,17,35]. Thus, connections between several methodologies previously used to analyze the topology of trees, such as size distributions [10,23], unbalance and depth [4,8,32–35], and transport efficiency [7,13,27,28], are revealed within the framework presented here.

## Supporting Information

**Text S1** Scaling of branch size and cumulative branch size: TreeBASE vs. manually selected data sets. We provide the list of references corresponding to the selected intraspecific and interspecific phylogenetic trees; the statistics of all data sets with two specific examples; and a summary table of taxa in the data sets.

Found at: doi:10.1371/journal.pone.0002757.s001 (0.06 MB DOC)

**Table S1** Break-down of the number of analyzed inter- and intra-species trees with respect to taxa.
Found at: doi:10.1371/journal.pone.0002757.s002 (0.03 MB DOC)

**Figure S1** Cumulative complementary distribution functions (CCDFs) for branch size ($F(A)$, panel A) and cumulative branch size ($F(C)$, panel B), and the allometric scaling relation ($C$ {similar, tilde operator} $A^n$, panel B) averaged and logarithmically binned over all phylogenetic trees. Empty squares are for the interspecific TreeBASE data set, solid circles are for the manually compiled intraspecific data set, and triangles are for the new manually compiled interspecific data set of reduced size. Solid lines are power laws fitted to the TreeBASE behavior, as in Figs. 2 and 3 of the main text.
Found at: doi:10.1371/journal.pone.0002757.s003 (1.22 MB TIF)

**Figure S2** A: An example of an intraspecific phylogenetic tree: different strains of the bacteria Vibrio vulnificus [S19]. Most of the branchings are binary, but there are some 3rd order branchings. B: The allometric scaling plot showing the relationship of cumulative branch size ($C$) to branch size ($A$) from each node of that tree. The solid line corresponds to the fitting $C$ {similar, tilde operator} $A^{1.43}$ to this intraspecific dataset.
Found at: doi:10.1371/journal.pone.0002757.s004 (2.66 MB TIF)

**Figure S3** A: An example of an interspecific phylogenetic tree: the catfish species (order Siluriformes) [S80]. Most of the branchings are binary, but there are some 3rd order branchings. B: The allometric scaling plot showing the relationship of cumulative branch size ($C$) to branch size ($A$) from each node of that tree. The solid line corresponds to the fitting $C$ {similar, tilde operator} $A^{1.44}$ to this intraspecific dataset.
Found at: doi:10.1371/journal.pone.0002757.s005 (2.58 MB TIF)

## Author Contributions

Conceived and designed the experiments: VME. Analyzed the data: EAH CJT KK VME. Contributed reagents/materials/analysis tools: VME. Wrote the paper: EAH VME EHG CMD. Discussed and interpreted the results: CD EH KK CT EH.

# SUPPORTING INFORMATION

*Scaling of branch size and cumulative branch size: TreeBASE vs. manually selected data sets*

The interspecific data set analyzed in this paper consists of 5212 phylogenetic trees downloaded from TreeBASE (`http://www.treebase.org`). Given that a database similar to TreeBASE does not exist for intraspecific phylogenies, we constructed our intraspecific data set by manually compiling 67 phylogenetic trees from several published references [S1-S45]. The difference in size between the two data sets calls for some additional checking on the appropriateness of a comparison between them. As a way to close the gap between the two datasets we compiled a third set of trees consisting of phylogenies of interspecific character, like the data in TreeBASE, but manually extracted from published references [S46-85] following the same criteria as the intraspecific set analyzed in the paper, and with the same size, 67 trees. We remind (see main text) that our selection criteria insure that our tree datasets contained organisms from terrestrial, marine and fresh water environments, from all the main climatic regions, from all kingdoms (Table S1), and reconstructed with the main phylogenetic tree estimation methods.

The results of this analysis are shown in Figure S1 (we illustrate tree structures and the allometric scaling for one intraspecific and one interspecific tree in Figures S2 and S3, respectively). It displays the cumulative complementary distribution functions (CCDFs) for branch size ($F(A)$, panel a) and cumulative branch size ($F(C)$, panel b), and the allometric scaling relation ($C \sim A^\eta$, panel c) averaged and logarithmically binned over all phylogenetic trees. We see that, despite their different size, the two interspecific data sets display the same behavior. Any bias in the manual selection procedure with respect to TreeBASE, if present, is weak enough to have no impact on the topological scaling behavior. In addition, there is perfect agreement between the scaling of the three data sets, except for the largest tree sizes for which there is poor statistics in the smaller data sets. This gives further support to the universality of the scaling found.

**Table S1. Break-down of the number of analyzed inter- and intra-species trees with respect to taxa.**

**Figure S1. Scaling relations from the enlarged data set described in Supporting Information.** Cumulative complementary distribution functions (CCDFs) for branch size ($F(A)$, panel A) and cumulative branch size ($F(C)$, panel B), and the allometric scaling relation ($C \sim A^\eta$, panel C) averaged and logarithmically binned over all phylogenetic trees. Empty squares are for the interspecific TreeBASE data set, solid circles are for the manually compiled intraspecific data set, and triangles are for the new manually compiled interspecific data set of reduced size. Solid lines are power laws fitted to the TreeBASE behavior, as in Figs. 2 and 3 of the main text.

**Figure S2. Intraspecific phylogenetic tree.** A: An example of an intraspecific phylogenetic tree: different strains of the bacteria *Vibrio vulnificus* [S19]. Most of the branchings are binary, but there are some 3rd order branchings. B: The allometric scaling plot showing the relationship of cumulative branch size ($C$) to branch size ($A$) from each node of that tree. The solid line corresponds to the fitting $C \sim A^{1.43}$ to this intraspecific dataset.

**Figure S3. Interspecific phylogenetic tree.** A: An example of an interspecific phylogenetic tree: the catfish species (order *Siluriformes*) [S80]. Most of the branchings are binary, but there are some 3rd order branchings. B: The allometric scaling plot showing the relationship of cumulative branch size (*C*) to branch size (*A*) from each node of that tree. The solid line corresponds to the fitting $C \sim A^{1.44}$ to this intraspecific dataset.

|  | INTER | INTRA |
|---|---|---|
| *Animalia* | 26 | 24 |
| *Archaea* | 3 | 0 |
| *Bacteria* | 9 | 18 |
| *Fungi* | 13 | 6 |
| *Plantae* | 8 | 6 |
| *Protozoa* | 6 | 4 |
| *Viruses* | 2 | 9 |

**Table S1. Break-down of the number of analyzed inter- and intra-species trees with respect to taxa.**

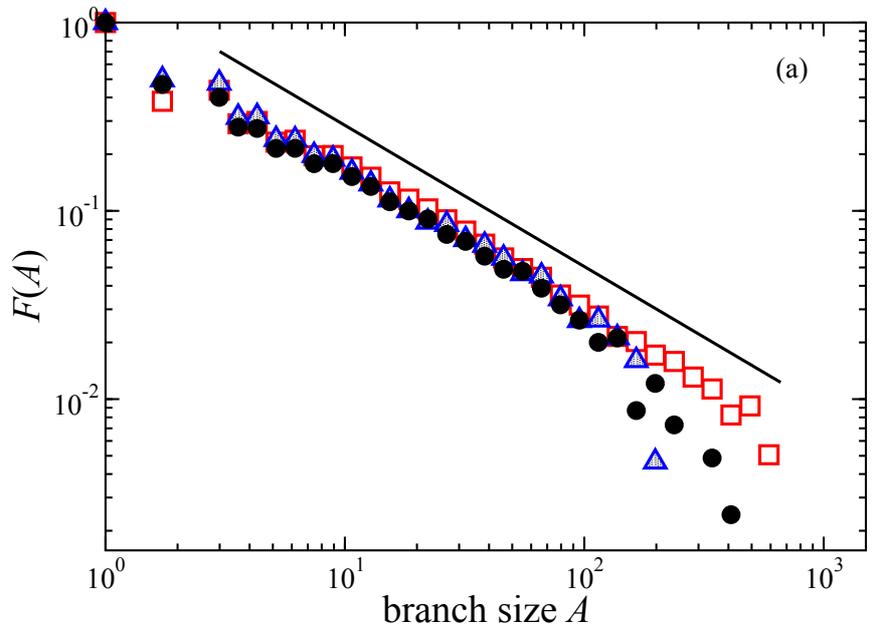

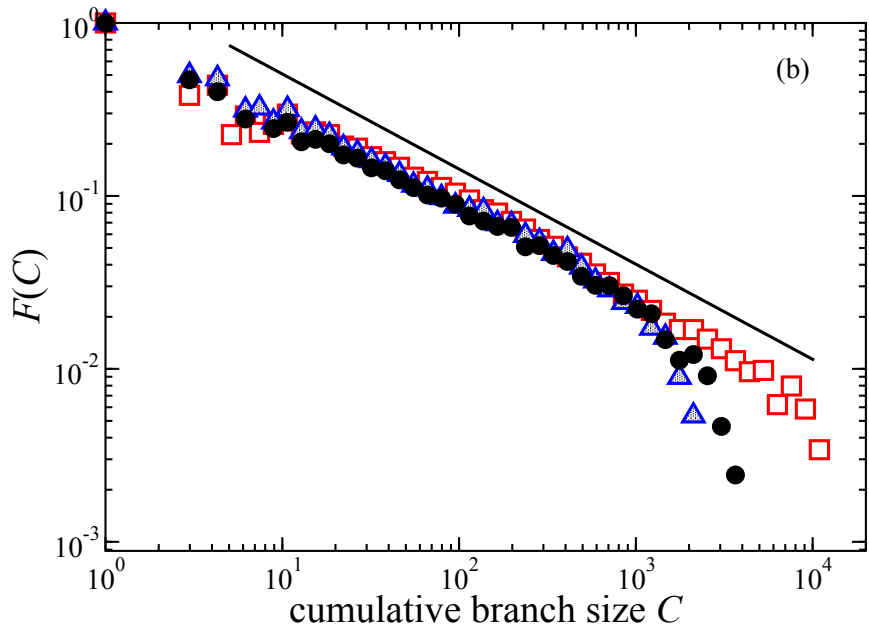

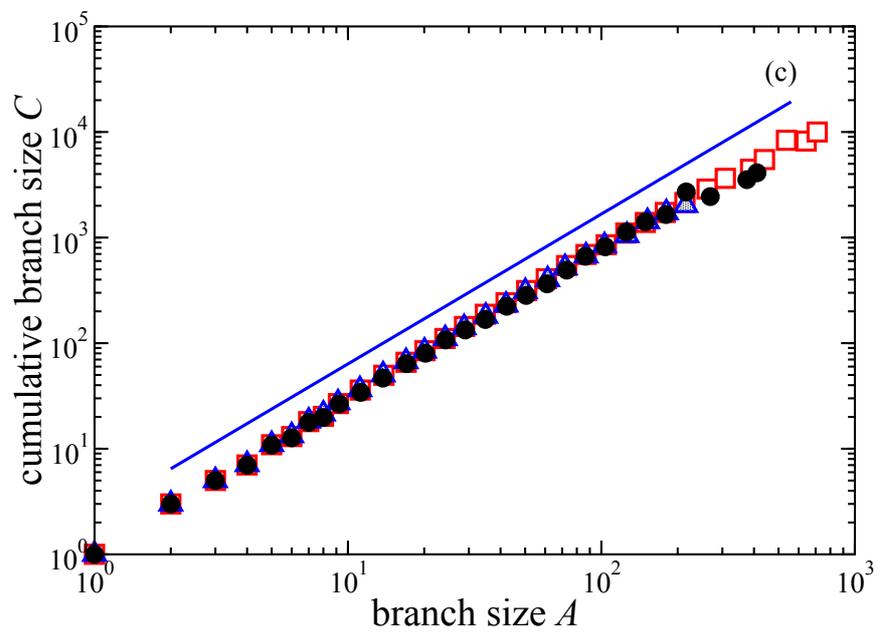

FIGURE S1

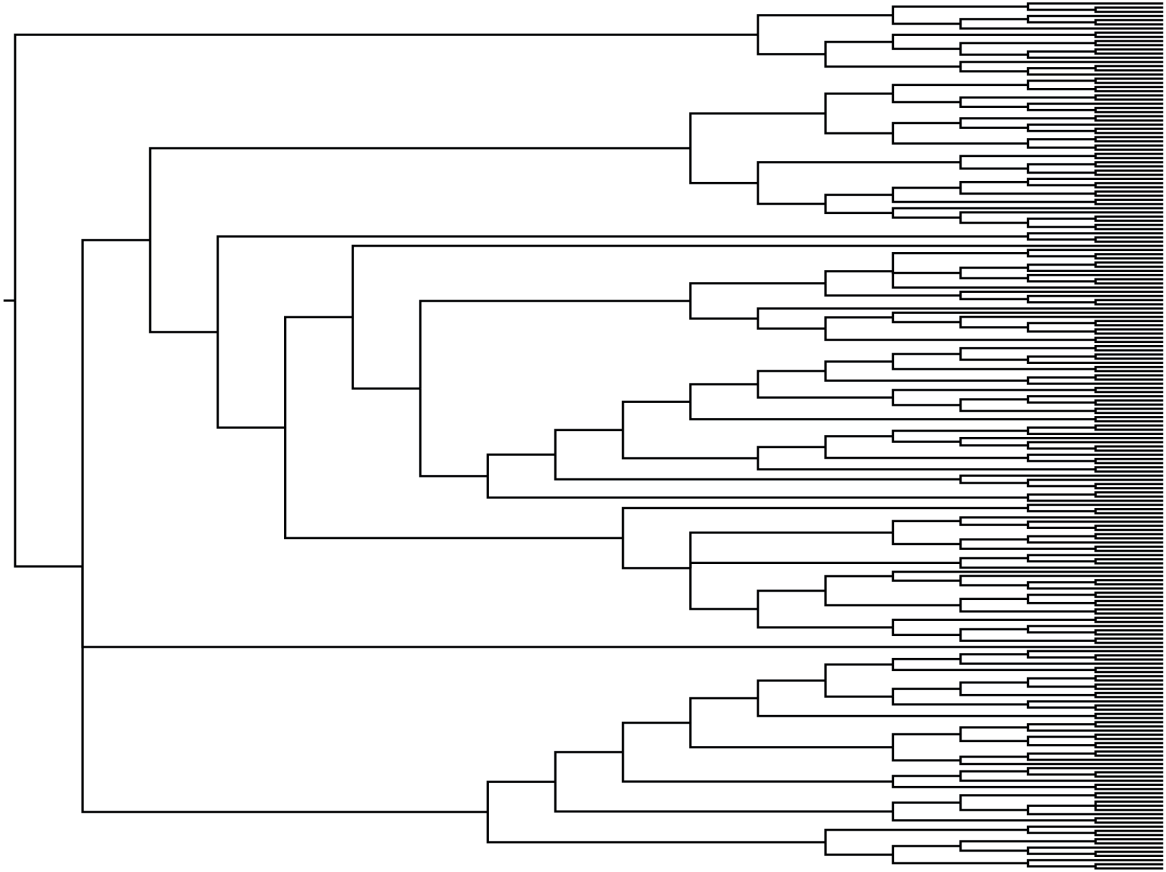

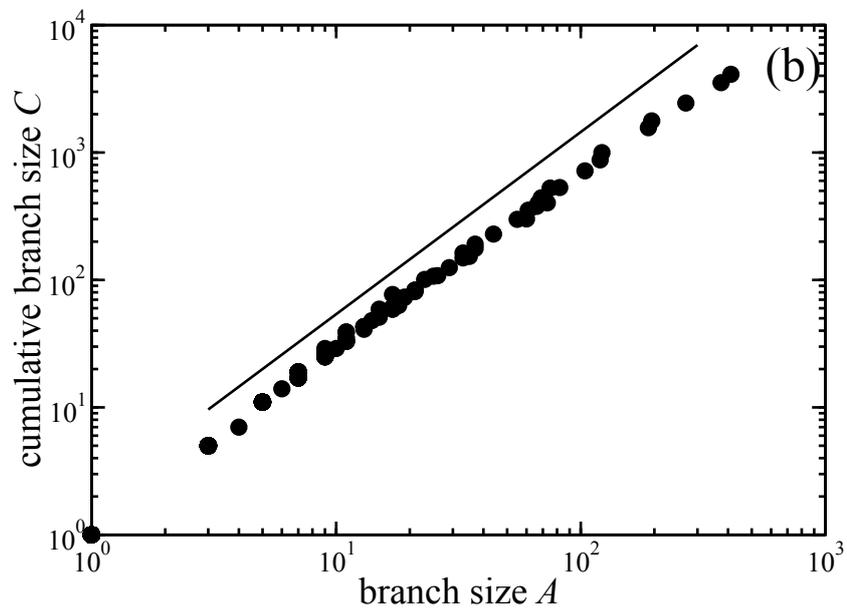

FIGURE S2

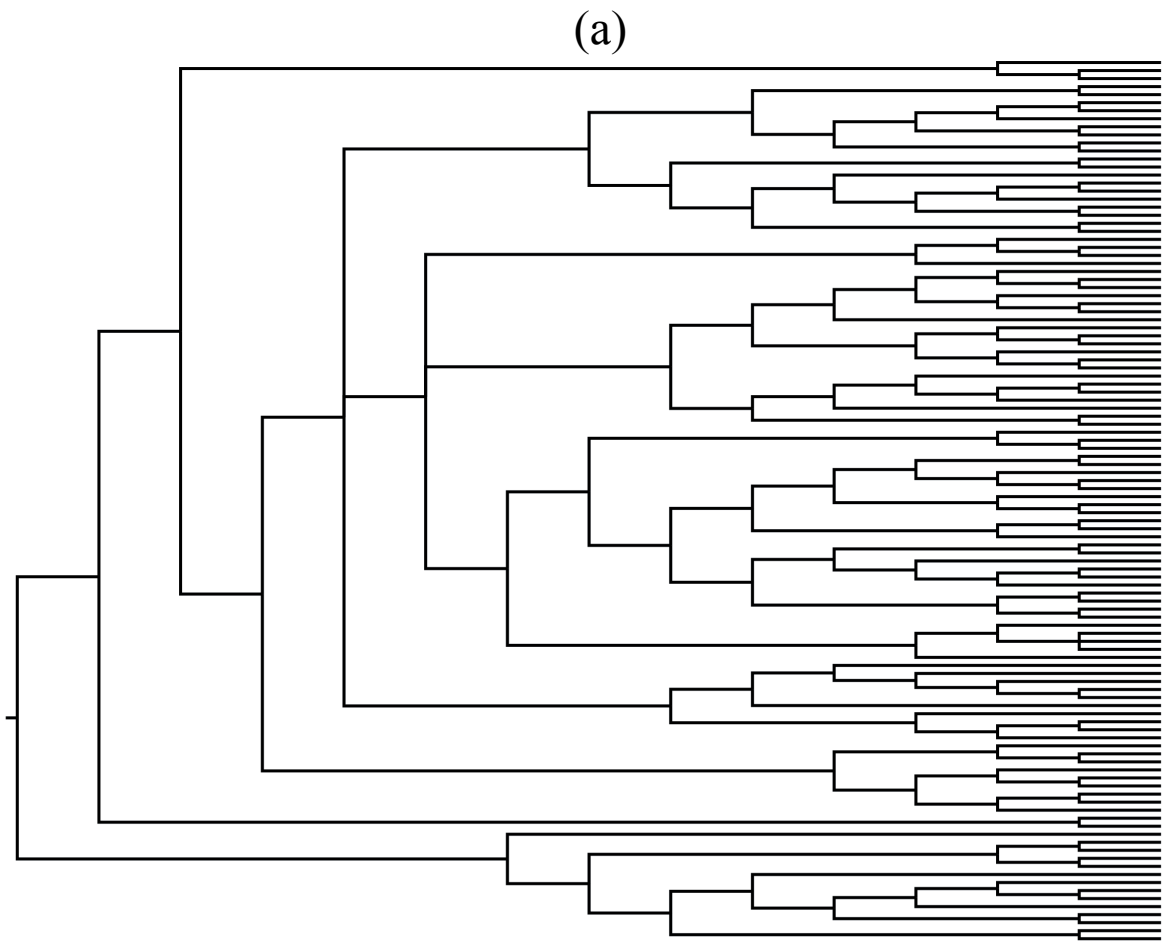

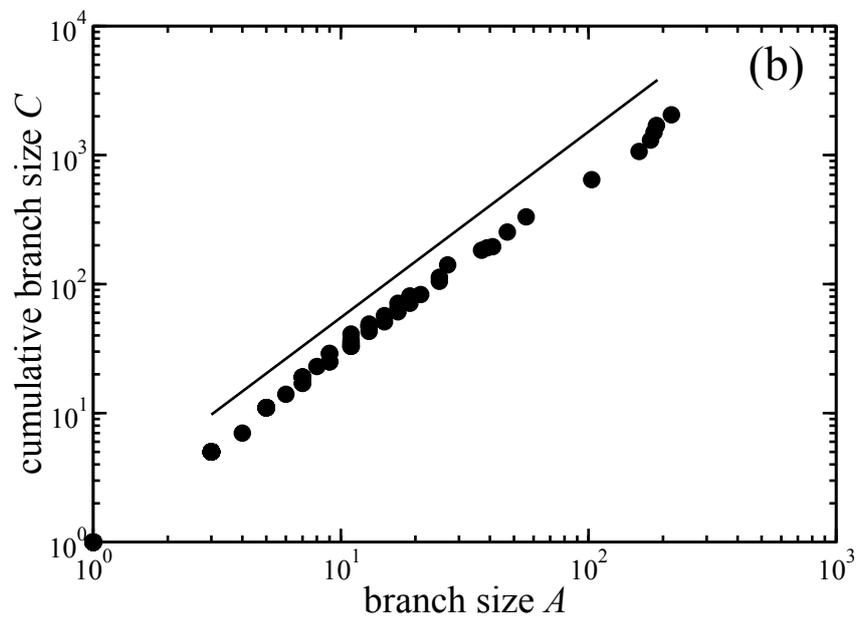

FIGURE S3

**Intraspecific and interspecific data sets**

The intraspecific and interspecific phylogenies which have been analyzed, in addition to the ones from TreeBase, have been obtained from the following references:

*Intraspecific phylogenies*

*Interspecific phylogenies*